\documentclass[11pt,onecolumn,dvips,draftcls]{IEEEtran}

\usepackage{amsfonts,amsmath,graphicx,epsfig}
\usepackage{subfigure,graphics,amssymb,amsxtra,color}
\usepackage{algorithm}
\usepackage{algorithmic}
\usepackage[ansinew]{inputenc}

\newtheorem{theorem}{Theorem}
\newcommand{\prob}{{\cal P}}

\newcommand{\avg}[2]{{\langle #1 \rangle}_{#2}}
\graphicspath{{graphics/}}

\begin{document}

\title{Wireless Information-Theoretic Security
--- Part II: Practical Implementation
\footnote{
Matthieu Bloch and Steven W. McLaughlin are with GT-CNRS UMI 2958, Metz, France, and also with the School of ECE, Georgia Institute of Technology, Altlanta, GA.\\
\indent Jo\~ao Barros is with the Departament of Computer Science \& LIACC/UP,
Universidade do Porto,  Portugal.\\
\indent Miguel R. D. Rodrigues is with the Computer Laboratory, University of Cambridge, United Kingdom.\\
\indent Parts of this work have been presented at the IEEE International Symposium on Information Theory 2006~\cite{Bloch2006a}, at the 44th Allerton conference on Communication Control and Computing~\cite{Bloch2006b}, and at the IEEE Information Theory Wokshop 2006 in Chengdu~\cite{Bloch2006c}.}
}
\author{\authorblockN{Matthieu Bloch
, Jo\~ao Barros
, Miguel R. D. Rodrigues
, and Steven W. McLaughlin
}
} \maketitle

\begin{abstract}
In Part I of this two-part paper on confidential communication over wireless channels, we studied the fundamental security limits of
quasi-static fading channels from the point of view of outage secrecy capacity with perfect and imperfect channel state information.
In Part II, we develop a practical secret key agreement protocol for Gaussian and quasi-static fading wiretap channels.  The protocol uses a four-step procedure to secure communications:  establish common randomness via an opportunistic transmission, perform message reconciliation, establish a common key via privacy amplification, and use of the key. We introduce a new reconciliation procedure that uses multilevel coding and optimized low density parity check codes which in some cases comes close to achieving the secrecy capacity limits established in Part I. Finally, we develop new metrics for assessing average secure key generation rates and show that our protocol is effective in secure key renewal.
\end{abstract}

\newpage

\section{Introduction}

Part I of this two-part paper was devoted to the information-theoretic security limits of a wireless communications
scenario with quasi-static fading. The analysis was carried out in terms of outage probability and outage secrecy
capacity, both for perfect and imperfect channel state information. In the second of this two-part paper we discuss the practical aspects associated with coding and modulation for the Gaussian and quasi-static fading wiretap channels. Virtually all systems today separate the problems of reliable and secure communications and provides them in a tandem fashion, where reliable communication is dealt with at the physical layer and security is provided at a higher layer (e.g. the network, transport or application layers) after the physical layer has been established. In this paper we show how modern physical layer tools, such as modulation, multilevel coding (MLC) and error control codes can be combined with key agreement protocols to, in some cases, come close to the fundamental limits described in Part I of this paper.

The general problem of physical layer-based coding and modulation schemes for both reliable and secure communication over Gaussian and fading wiretap channels has not received much attention and there is no larger framework to draw on, even with the sustained advances in the area coding and modulation for Gaussian and fading channels \cite{Forney1998,Biglieri1998}. Much of previous  work for the wiretap channel stems from the early work~\cite{Wyner1975} and~\cite{Ozarow1984} and studied more extensively by Wei~\cite{Wei1991}. This work shows how to encode secret information using cosets of certain linear block codes.  More recently, this general notion has been extended by Thangaraj {\em et al.}~\cite{Thangaraj2006} where it was shown how low density parity check codes can asymptotically achieve the secrecy capacity for the erasure wiretap channel, and how it can be used to provide perfectly secret communications at rates below the secrecy capacity for other channels. Thangaraj et al.~\cite{Thangaraj2006} also showed how the joint problems of reliability and security interact in a code and how capacity approaching codes for the reliability problem can be used for reliability and security requirements of the wiretap channel. Existence of coding schemes for various generalized wiretap channel scenarios has been proved by several authors recently~\cite{Hayashi2004,Hayashi2006,Muramatsu2006}. In particular, the existence of coding methods based on LDPC codes has been shown in~\cite{Muramatsu2006}.


Since designing wiretap codes for Gaussian and fading channels appears to be beyond the capabilities of current coding techniques,
we focus on the somewhat easier problem of generating secret keys for secure communication over wireless channels.
The key generation/distribution problem in wiretap channels falls under the general problem of key generation from correlated source outputs, which has been studied~\cite{Maurer1993,Maurer1999,Ahlswede1993} in an information theoretic context. The objective of secure key distribution is for Alice and Bob to agree on a common $k$-bit key about which Eve's entropy is maximal. In key distribution, the $k$ bits can be unknown to Alice before transmission, this is in contrast to secure message communication where Alice has a $k$-bit message that she wants to communicate to Bob, we focus only on the former. Powerful tools such as common randomness, advantage distillation and privacy amplification were developed in the context of key distribution over wiretap channels (~\cite{Ahlswede1993,Bennett1995}) and will be discussed, as they form the basis for much of the practical secret key agreement protocol proposed in this paper.  Most of the key agreement protocols require some level of interactive communication between Alice and Bob to arrive at a common but secret key~\cite{Maurer1993}, where they exchange information by way of a parallel, error-free public channel between Alice and Bob during the key agreement phase (e.g.~\cite{VanAssche2004}).  One key advance in this paper is that we focus exclusively on protocols that require only one-way, feed-forward communication from Alice to Bob across the noisy wireless channel and there is no need for a noiseless, authenticated public channel.  


\subsection{Our Contributions}  

Our main contributions are as follows:

{$\bullet$} Development of a secret key agreement protocol for the Gaussian channel that performs close to the fundamental secrecy capacity limits
(determined in Part I) over a wide range of channel values.  The communication is from the transmitter-to-receiver only and requires no feedback or error-free side channels.

{$\bullet$} Adaptation of the secret key agreement protocol for the Gaussian channel to the quasi-static fading channel with perfect channel state information. In some cases this protocol comes close to the fundamental limits of the wireless fading channels presented in Part I. Again, the communication is from the transmitter-to-receiver only and requires no feedback or error-free side channels.

{$\bullet$} Extension of the secret key agreement protocol for the quasi-static fading channel to the case of imperfect channel information.

{$\bullet$} Development of new security and communication metrics, such as {\em average $\eta$-secure throughput} and {\em average $\eta$-communication throughput}  for average secret and non- secret bits, respectively, transmitted per channel use on the wiretap channel.

\subsection{Organization of the Paper}

The remainder of the paper is organized as follows. In Section \ref{sec:Funda} we consider a one-way protocol for key agreement for the Gaussian channel.  
In Section \ref{sec:reconciliation} we give a new reconciliation procedure for the Gaussian channel that is based on multilevel coding and LDPC codes. In Section \ref{sec:opportunSec} we extend this protocol in an {\em opportunistic} way to the quasi-static fading channel and show that in some instances (when both the main and wiretapper's channel have low SNR) that the protocol comes very close to the secrecy capacity.
  We also show how the effect of imperfect channel knowledge on the performance of the protocol.  Finally we provide concluding remarks and next steps.

\section{Secret Key Agreement over Gaussian Channels}

\label{sec:Funda}

\begin{figure}[htbp]

     \centering

          \includegraphics[width=10cm]{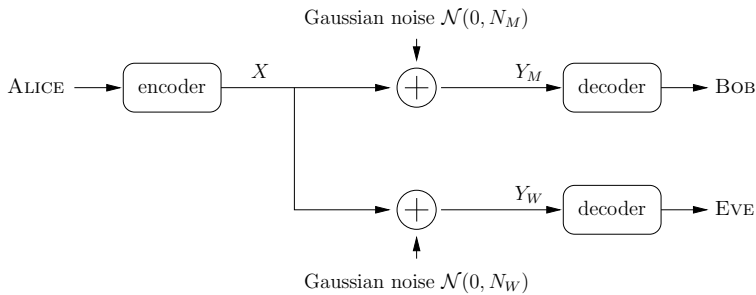}

     \caption{The Gaussian wiretap channel.}

     \label{fig:GaussChan}

\end{figure}

As a prelude to the problem of coding for the quasi-static fading wiretap channel, we develop a protocol for the Gaussian wiretap channel shown in Figure~\ref{fig:GaussChan}.

It is assumed that both channels are discrete time additive Gaussian noise channels with an average transmitted power constraint of 1 and the noise on Eve's channel is independent of the noise on the main channel between Alice and Bob. The noise variances for the main and wiretap channel are denoted respectively $N_M$ and $N_W$. Furthermore we assume that Eve's wiretapper's channel is {worse} than the main channel, namely $N_W > N_M$.  This critical assumption is necessary to ensure that the secrecy capacity is strictly positive~\cite{Leung-Yan-Cheong1978}.  If noiseless feedback is allowed between Bob and Alice, then $N_M > N_W$ is permitted and the secrecy capacity can be positive~\cite{Maurer1993}, however we consider only the case of one-way, noisy communications from Alice to Bob and $N_W > N_M$.  There are a number of practical scenarios where this assumption is valid, for example radio frequency identification (RFID) tags and readers with a passive eavesdropper~\cite{Chabanne2006}.

\begin{figure}[htbp]

     \centering

          \includegraphics[width=9cm]{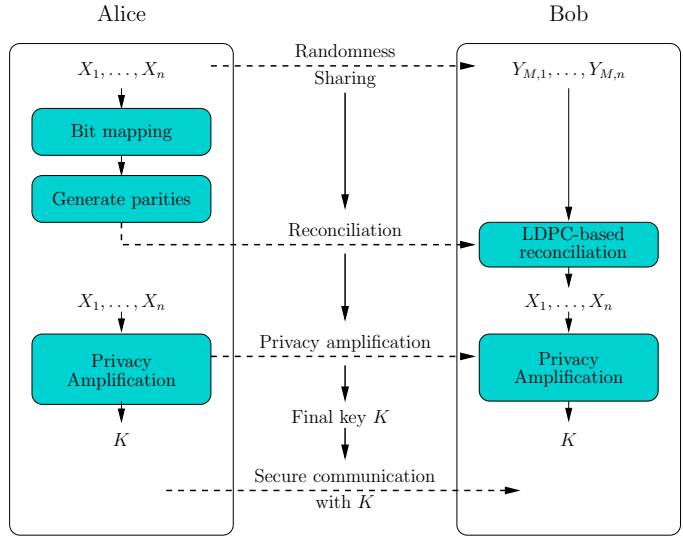}

     \caption{The four-step procedure for secret key agreement on the Gaussian channel.}

     \label{fig:GaussProtocol}

\end{figure}

Although the secrecy capacity of the Gaussian wiretap channel has been fully characterized~\cite{Leung-Yan-Cheong1978}, designing practical coding schemes is still an open problem. On the other hand, previous results on secret key agreement by public discussion~\cite{Maurer1993} and privacy amplification~\cite{Bennett1995} naturally suggest a four step approach to secure communication over a wiretap channel:  {\em randomness sharing, information reconciliation, privacy amplification, secure communication}.  In this section we show how to adapt these to the Gaussian channel. The following protocol exploits a more general and more efficient version of the information reconciliation method of ~\cite{VanAssche2004} and is shown in Figure~\ref{fig:GaussProtocol}.  Some of the four steps are adapted directly from previous work and not modified, while others require further development as given later in the paper.

\textbf{1.  Randomness sharing}. The existence of common information between Alice and Bob is the key ingredient required for secret key agreement. In a wiretap scenario, Alice can generate this shared randomness by transmitting a sequence $X^n =\left(X_1,\dots,X_n\right)$ of $n$ i.i.d. realizations of a discrete random variable $X$ over the main channel, which will provide Bob and Eve with sequences of correlated continuous variables $Y_M^n = \left(Y_{M,1},\dots,Y_{M,n} \right)$ and $Y_W^n = \left(Y_{W,1},\dots,Y_{W,n} \right)$ respectively.  In Figure~\ref{fig:GaussProtocol} the dotted lines indicate the transmission across the Gaussian channel to Bob (and Eve, not shown here).

Since the amount of secrecy extractable from this common randomness is known to be at least~\cite{Maurer1993}

\begin{equation}
S_{min}\geq I(X;Y_M)-\min\left(I(X;Y_W),I(Y_M;Y_W)\right)\;\mbox{bits/symbols},
\end{equation}

the mutual information $I(X;Y_M)$ should be maximized and Alice should therefore choose $X$ achieving the capacity $C_M=0.5\log_2(1+1/N_M)$ of the main channel. Matching $C_M$ exactly is only possible with continuous Gaussian random variables, however the set $\mathcal{X}$ and the probability mass function of $X$ can be optimized so that $I(X;Y_M)$ lies within hundredth of bits of the channel capacity $C_M$ even with a discrete distribution. For a fixed number of constellation points $N_c=\left|\mathcal{X}\right|$, this optimization can be performed with the algorithm proposed in~\cite{Varnica2002}, however a very good approximation of the optimum can simply be obtained by expanding a uniformly spaced amplitude shift keying (ASK) constellation $\left\{x_i\right\}_{i=1\dots N_c}=\left\{\pm 1,\pm 3,\dots,\pm\frac{N_c-1}{2}\right\}$ by a factor $\alpha\in\mathbb{R}^+$, and using a Maxwell-Boltzmann probability distribution

\begin{equation}
\label{eq:Maxwell}
P(X=x_i) = \frac{\exp{\left(-\lambda \alpha^2{\left|x_i\right|}^2\right)}}{\sum_j \exp{\left(-\lambda \alpha^2{\left|x_i\right|}^2\right)}}.
\end{equation}

Even though $I(X;Y_M)$ is not a convex function of $\alpha$ and $\lambda$, non-linear programming seems to be relatively insensitive to the initialization of the optimization.  Clearly, $N_c$ should be large enough so that $I(X;Y_M)$ can approach $C_M$ within the required precision, its exact choice will be discussed in the Section~\ref{sec:reconciliation}.









\textbf{2. Information reconciliation}~\cite{Brassard1993}. The channel noise introduces discrepancies between Bob's received symbols $Y_M^n$ and Alice's symbols $X^n$. The first step is for Bob to estimate Alice's symbols $\hat{X}^n=\left( \hat{X}_{1},\dots,\hat{X}_{n} \right)$ based on $Y_M^n$. The channel noise results in discrepancies in the correlated bit sequences $X^n$ and $\hat{X}^n$ that Alice and Bob will correct and reconcile before any further processing. This requires an additional exchange of information between Alice and Bob as shown in Figure~\ref{fig:GaussProtocol}, which is also made available to Eve. This situation can be viewed as a special case of source coding with side-information, where Alice compresses her source $X^n$ and Bob decodes it with the help of correlated side information $Y_M^n$. The Slepian-Wolf theorem~\cite{Slepian1973} yields a lower bound on the total number of bits $M_{rec}$ which have to be exchanged:

\begin{equation}
M_{rec}\geq H(X^n|Y_M^n) = n H(X|Y_M).
\end{equation}

Notice that the result of~\cite{Slepian1973} only applies to discrete random variables whereas here $Y_M$ is continuous. The variable $Y_M$ can however be quantized into a discrete random variable $Y_q$ such that $H(X|Y_q)$ approaches $H(X|Y_M)$ with arbitrary precision, and the Slepian-Wolf Theorem still holds.

Practical reconciliation algorithms will introduce an overhead $\epsilon_{rec}>0$ and require the transmission of $M_{rec}=nH(X|Y_M)(1+\epsilon_{rec})$ additional bits. The reconciliation can also be characterized by its efficiency $\beta$ which is defined as

\begin{equation}
\beta(\epsilon_{rec}) = 1-\epsilon_{rec}\frac{H(X|Y_M)}{I(X;Y_M)} \leq 1.
\end{equation}

At the end of the reconciliation step, Alice an Bob share with high probability the common sequence $X^n$ whose entropy is $n_{rec}=nH(X)$. We will assume that $X^n$ is then compressed into a $n_{rec}$-bits binary sequence $S$.  For our application to the Gaussian wiretap channel we use multilevel coding (at Alice) and multistage (MS) decoding (at Bob) to reconcile and correct the differences between $\hat{X}$ and $X$ and this is discussed in detail in Section~\ref{sec:reconciliation}.

\textbf{3. Privacy amplification}~\cite{Bennett1995}. This last operation allows Alice and Bob to extract a secret key from the binary sequence $S$. The principle of privacy amplification is to apply a well-chosen compression function $g:{\left\{0,1\right\}}^{n_{rec}}\rightarrow{\left\{0,1\right\}}^{k}$ ($k<n_{rec}$) to the reconciled bit sequence, such that the eavesdroppers obtains negligible information about the final $k$-bit sequence $g(S)$. In practice this can be achieved by choosing $g$ at random within a universal family of hash functions~\cite{Carter1979}, as stated in the following theorem.

\begin{theorem}{\cite[Corollary 4]{Bennett1995}}
\label{th:general_privacy_amp}
Let $S\in{\left\{0,1\right\}}^{n_{rec}}$ be the random variable representing the bit sequence shared by Alice and Bob, and let $E$ be the random variable representing the total information available to the eavesdropper. Let $e$ be a particular realization of $E$. If the R\'enyi entropy (of order 2) $R(S|E=e)$ is know to be at least $c$, and Alice and Bob choose $K=G(S)$ as their secret key, where $G$ is a hash function chosen at random from a universal family of hash functions $\mathcal{G}:{\left\{0,1\right\}}^{n_{rec}}\rightarrow{\left\{0,1\right\}}^{k}$, then
\begin{equation}
\label{eq:uncertainty_bound}
H(K|G,E=e)\geq k-\frac{2^{k-c}}{\ln 2}.
\end{equation}
\end{theorem}
The total information available to Eve $E$ consists in the sequence $Y_W^n$ received during the first stage of the protocol, as well as the additional bits echanged during reconciliation, represented by the random variable $M$. As shown in~\cite[Theorem 5.2]{Cachin1997} :
\begin{equation}
R(S|Y_W^n=y_w^n,M=m)\geq R(S|Y_W^n=y_w^n) - \log_2{\left|M\right|}-2s-2\quad\mbox{with probability}\quad 1-2^{-s}.
\end{equation}
The quantity $\log_2{\left|M\right|}$ represents the number of bits intercepted by Eve during the reconciliation, which is at most $nH(X|Y_m)(1+\epsilon_{rec})$ if she intercepted all the information. Evaluating $R(S|Y_W^n=y_W^n)$  is in general still difficult, however conditioned on the typicality of the bit sequence~\cite{Maurer2000} $R(S|Y_W^n=y_W^n)$ and $H(S|Y_W^n=y_W^n)$ become equal. Hence if $n$ is large enough, $nH(X|Y_W)-nH(X|Y_M)(1+\epsilon_{rec})-2s-2$ is a good lower bound of $R(S|E=e)$, and choosing
\begin{equation}
k = n\beta I(X;Y_M) - nI(X;Y_W)-2s-2-r_0,
\end{equation}
guarantees that Eve's uncertainty on the key is greater than $k-2^{-r_0}/\ln 2$ with probability $1-2^{-s}$. For our protocol in this paper we do not develop anything new, and we use standard families of hash functions~\cite{Carter1979,Wegman1981}.

\textbf{4. Secure communication}. The secret key generated $K=G(S)$ can finally be used to secure Alice's message, using either a one-time pad for perfect secrecy or a standard secret key encryption algorithm and Eve's uncertainty $H(K|G,E=e)$ about the key is as close to $k$ as we want as per~(\ref{eq:uncertainty_bound}).

\section{LDPC Constructions for Gaussian Reconciliation}

\label{sec:reconciliation}

In this section we develop an efficient reconciliation approach for Step 2 of the key agreement. The reconciliation of binary random variables has been extensively studied and several efficient methods have been proposed~\cite{Brassard1993,Buttler2003}, however little attention has been devoted to the practical reconciliation of non-binary random variables~\cite{VanAssche2004}. As stated previously the goal is given a non-binary variable $X$ with distribution given by Eq.~(\ref{eq:Maxwell}) and a random variable $Y_M$ obtained by sending $X$ through an additive Gaussian channel with noise variance $N_M$, to generate a minimum amount of (parity) information to be sent to Bob so that $X$ can be recovered from $Y_M$ and this additional information.

\subsection{Multilevel LDPC Codes for Slepian-Wolf Compression}

We assume here that Alice and Bob have, respectively, access to the outcomes $x^n = {\left\{x_i\right\}}_{i=0\dots n-1}\in\mathcal{X}^n$ and $y^n = {\left\{y_i\right\}}_{i=0\dots n-1}\in\mathbb{R}^n$ of instances of the random variables $X^N$ and $Y_M^n$. Alice sould then send Bob additional information to help him recover $x^n$ based on $y^n$, and we can assume without restriction that Bob recovers a binary description of $x^n$. Since each element of $\mathcal{X}$ can be uniquely described by a $m$-bit label ($m\geq\log_2\left|\mathcal{X}\right|$), we introduce the $m$ labeling functions $\ell_k:\mathcal{X}\rightarrow{\left\{0,1\right\}}$ ($k\in\left\{0\dots m-1\right\}$), which associate to any element of $\mathcal{X}$ the $k$th bit of its binary label. As suggested in~\cite{Wyner1974}, we can then use the syndromes of $\left\{\ell_k(x_i)\right\}_{k=0\dots m-1}^{i=0\dots n-1}$ according to a binary code as the additional information sent by Alice to Bob.

\begin{figure}[htbp]

     \centering

          \includegraphics[width=10cm]{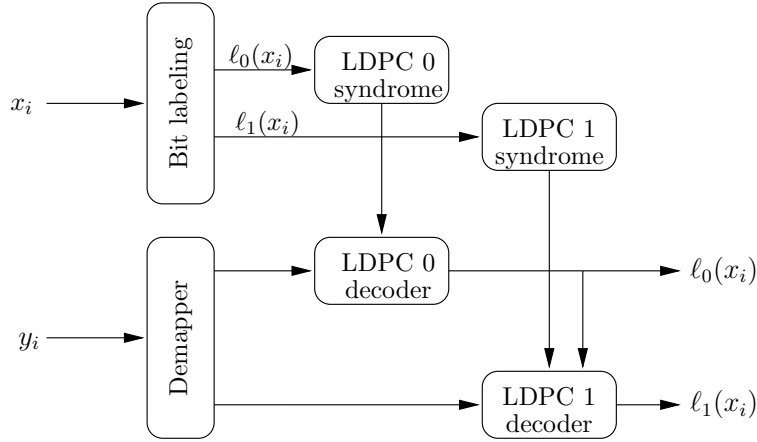}

     \caption{Principle of MLC/MSD reconciliation in the case m=2.}

     \label{fig:reconciliation}

\end{figure}

Because of the particular correlation considered here, the reconciliation of $X$ and $Y_M$ is similar to a coded modulation scheme, where Alice would transmit her data over a Gaussian channel using a Pulse-Amplitude-Modulation scheme. Most standard modulation techniques such as Bit Interleaved Coded Modulation (BICM)~\cite{Caire1998} or MultiLevel Coding/MultiStage Decoding (MLC/MSD)~\cite{Wachsmann1999} schemes can therefore be adapted to reconciliation. In the case of a BICM-like reconciliation, a single syndrome would be computed based on an interleaved version of the bit sequence $\left\{\ell_k(x_i)\right\}_{k=0\dots m-1}^{i=0\dots n-1}$, whereas in the case of MLC/MSD-like reconciliation, the $m$ syndromes of the sub-sequences $\left\{\ell_k(x_i)\right\}_{i=0\dots n-1}$ ($k\in\left\{0\dots m-1\right\}$) would be computed successively, as illustrated in Figure~\ref{fig:reconciliation}.

In what follows we will describe a reconciliation algorithm adapted from the last scheme. This choice was motivated by the fact that BICM is known to be suboptimal over the Gaussian channel, hence the reconciliation of the variables $X$ and $Y_M$ with a BICM-like scheme would always require strictly more that $H(X|Y_M)$ additional bits per symbol. Moreover MLC/MSD is based on several components codes and therefore offers more flexibility on the code design than BICM.

The proposed reconciliation algorithm is a MLC/MSD-like reconciliation that uses binary LDPC component codes. Other classes of codes such as Turbo-Codes could be used as well, however LDPC have already proved their good performance for error-correction and side information coding~\cite{Liveris2002}, and the Belief-Propagation algorithm can easily be generalized to account for the correlation between the sub-sequences $\left\{\ell_k(x_i)\right\}_{i=0\dots n-1}$ ($k\in\left\{0\dots m-1\right\}$). We use the following notations to describe the algorithm:

\begin{itemize}

\item $b_i^k = \ell_k(x_i)$ ($i\in\left\{0\dots n-1\right\}$,$k\in\left\{0\dots m-1\right\}$),

\item $c(k)$ represents the number of check nodes at the $k$th level ($c(k)$ depends on the rate $R^k$ of the code used at level $k$ and will be discussed in the next section),

\item $m_{ij}^{(k,l)}$ denotes the message from a variable node $v_i^k$ ($i\in\left\{0\dots n-1\right\}$) to a check node $c_j^k$ ($j\in\left\{0\dots c(k)-1\right\}$) of the $k$th level at the $l$th iteration , and similarly $m_{ji}^{(k,l)}$ denotes the message from a check node $c_j^k$ to a variable node $v_i^k$ of the $k$th level at the $l$th iteration,

\item $\mathcal{M}_i^k$ denotes the set of all check nodes connected to the variable node $v_i^k$ of the $k$th level, and $\mathcal{N}_j^k$ denotes the set of all variables nodes connected to the check node $c_j^k$ of the $k$th level,

\item $s(c)$ is the syndrome bit associated to a check node $c$.

\end{itemize}

The $m$ levels are then decoded successively, and the update equations of the messages at the $l$th iteration of the belief propagation at a given level $k$ are described below :

\begin{eqnarray}
\forall i\in\left\{0\dots n-1\right\},\;\forall j\in\mathcal{M}_i^k\quad m_{ij}^{(k,l)} &=& \left\{\begin{array}{l} m_{i}^{(k,0)}\quad\mbox{if}\quad l=0,\\ m_{i}^{(k,0)}+\displaystyle\sum_{c_p\in\mathcal{M}_i^k\smallsetminus c_j}{\left(1-2s(c_p)\right)}m_{pi}^{(k,l)}\end{array}\right.,\label{eq:vn}\\
\forall i\in\left\{0\dots n-1\right\},\;\forall j\in\mathcal{M}_i^k\quad \lambda_{i}^{(k,l)} &=& \left\{\begin{array}{l} m_{i}^{(k,0)}\quad\mbox{if}\quad l=0,\\ m_{i,j}^{(k,l)}+\displaystyle{\left(1-2s(c_j)\right)}m_{ji}^{(k,l)}\end{array}\right.,\\
\forall j\in\left\{0\dots c(k)-1\right\},\;\forall i\in\mathcal{N}_j^k\quad m_{ji}^{(k,l)} &=& \log\frac{1+\displaystyle\prod_{v_p\in\mathcal{N}_j^k\smallsetminus v_i}\tanh\frac{m_{pj}^{(k,l-1)}}{2}}{1-\displaystyle\prod_{v_p\in\mathcal{N}_j^k\smallsetminus v_i}\tanh\frac{m_{p,j}^{(k,l-1)}}{2}},\label{eq:cn}
\end{eqnarray}
where
\begin{equation}
m_{i}^{(k,0)} = \log\frac{\displaystyle\sum_{\hat{x}\in\mathcal{X}:\ell_k(\hat{x})=1}p(y_i,\hat{x})\exp\left[-\sum_{p\neq k}(1-\ell_p(\hat{x}))(\lambda_i^{(p,l_{max})}-m_{i}^{(p,0)})\right]}{\displaystyle\sum_{\hat{x}\in\mathcal{X}:\ell_k(\hat{x})=0     }p(y_i,\hat{x})\exp\left[-\sum_{p\neq k}(1-\ell_p(\hat{x}))(\lambda_i^{(p,l_{max})}-m_{i}^{(p,0)})\right]}.\label{eq:interlevel}
\end{equation}
If the Tanner graphs of the LDPC component codes are trees, it can be shown that the values $\lambda_i^{(k,l)}$ converge to the true \textit{a posteriori} probabilities :
\begin{equation}
\frac{\prob\left[b_i^k=1|b_i^0\dots b_{i}^{k-1},y_i\right]}{\prob\left[b_i^k=0|b_i^0\dots, b_{i}^{k-1},y_i\right]}
\end{equation}
in a finite number of iterations, and the decision on the value of $b_i^k$ can finally made based on the sign of $\lambda_i^{(k,l_{max})}$. In practice, even when the Tanner graphs contain cycles, this belief-propagation algorithm still performs well.

The only difference between Eq.~(\ref{eq:vn})-(\ref{eq:cn}) and the standard update rules belief propagation is the term $m_i^{(k,0)}$, which takes into account not only the intrinsic information available from the observation $y_i$, but also from the decoding of the other levels $p\neq k$. Eq.~(\ref{eq:interlevel}) is similar to the update rule of a single-input single-output (SISO) demodulator, however it should be noted that it involves the joint probability $p(y,\hat{x})$ (and not the conditional probability $p(y|\hat{x})$) to take into account the non-uniform distribution of the symbols in $\mathcal{X}$. In theory, it should be sufficient to decode each level only once, however in practice performing several iterations between the levels might help improve the performance of the overall scheme. These practical issues will be discussed in the next section. Let us finally point out that the algorithms described in~\cite{Nana2006,Chen2006,Liveris2002} can all be viewed as special cases of this general algorithm.


\subsection{Rate Assignment}

\begin{figure}[htbp]

     \centering

          \includegraphics[width=10cm]{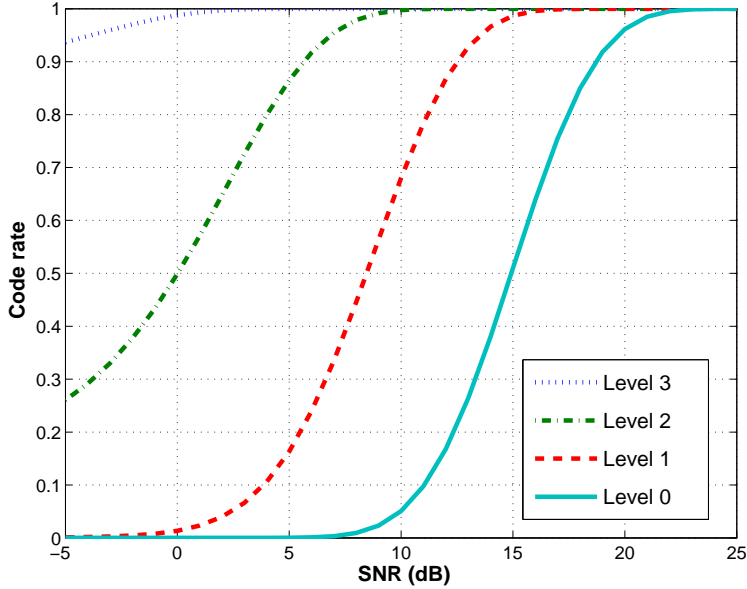}

     \caption{Optimal code rates required for the constellation of table~\ref{tab:refconst}}

     \label{fig:CodeRatesOpt}

\end{figure}

The optimal code rates required for each sub-sequence $\left\{\ell_k(x_i)\right\}_{i=0\dots n-1}$ are those required for MultiStage Decoding. In fact, from the chain rule of entropy we have :

\begin{equation}
H(X|Y_M) = H(\ell_0(X),\dots,\ell_{m-1}(X)|Y_M) = \sum_{k}H(\ell_k(X)|\ell_0(X),\dots,\ell_{k-1}(X),Y_M).\label{eq:crentrop}
\end{equation}
Hence the $H(X|Y_M)$ bits per symbol required for reconciliation can be obtained by disclosing successively $H(\ell_k(X)|\ell_0(X),\dots,\ell_{k-1}(X),Y_M)$ bits per symbol. The optimal code rate required at each level $k$ is therefore :
\begin{eqnarray}
R_{opt}^k &=& 1 - H(\ell_k(X)|\ell_0(X),\dots,\ell_{k-1}(X),Y_M).
\end{eqnarray}
Eq.~(\ref{eq:crentrop}) guarantees the optimality of the reconciliation scheme for any labeling, however the practical efficiency of the reconciliation strongly depends on the mapping used. In fact the performance of the reconciliation relies on our ability to construct capacity approaching codes for all levels $k$, which might not be possible if the required rates are too low. We investigated several labeling strategies and found out that the natural binary mapping was the best compromise. This mapping assigns to each symbol $\hat{x}_j\in\mathcal{X}$ the $m$-bit representation of $j+(2^m-\left|\mathcal{X}\right|)/2$, and $\ell_k(\hat{x}_j)$ is then the $k$th label bit ($\ell_0(\hat{x}_j)$ is the least significant bit). Figure~\ref{fig:CodeRatesOpt} shows the rates required for a constellation of size 10, with symbols and probabilities given in table~\ref{tab:refconst}, as a function of the signal to noise ratio $10\log_{10}(1/N_M)$.
\begin{table}[htbp]
     \centering
     \caption{Constellation optimized to maximize $I(X;Y_M)$ at a SNR of 13~dB.}
          \begin{tabular}{|c||c|c|c|c|c|c|c|c|c|c|c|c|}
               \hline
               $\hat{x}$ & -2.836&-2.320&-1.804&-1.289&-0.773&-0.258&0.258&0.773&1.289&1.804&2.320&2.836\\
               $\prob(\hat{x})$ & 0.0040&0.0146&0.0414&0.0904&0.1522&0.1974&0.1974&0.1522&0.0904&0.0414&0.0146&0.0040\\
               $\ell_0(\hat{x})$ & 0 &1 &0 &1 &0 &1 &0 &1 &0 &1&0 &1\\
               $\ell_1(\hat{x})$ & 1 &1 &0 &0 &1 &1 &0 &0 &1 &1&0 &0\\
               $\ell_2(\hat{x})$ & 0 &0 &1 &1 &1 &1 &0 &0 &0 &0 &1 &1\\
               $\ell_3(\hat{x})$ & 0 &0 &0 &0 &0 &0 &1 &1 &1 &1 &1 &1\\
               \hline
          \end{tabular}
     \label{tab:refconst}
\end{table}

Over a wide range of SNRs, the optimal rates of the two uppermost levels are equal to 1, which greatly simplifies code design by effectively requiring only two codes. We carried out extensive simulations, and observed that for any value of the SNR, adjusting the constellations size $N_c$ so that $H(X)\approx 0.5\log_2(1+SNR)$+1 would require at most two codes while still maintaining $I(X;Y_M)$ within a few hundredth of bits of its maximum value.

The natural mapping has the property of preserving the symmetry on the probability distribution of the random variable $X$:
\begin{equation}
\forall k\in\left\{0,\dots,m-1\right\},\forall y\in\mathbb{R},\forall\hat{x}_j\in\mathcal{X}\quad p(y,\ell_k(\hat{x}_j)) = p(-y,{\ell_k(\hat{x}_j)}\oplus 1).
\end{equation}
When first decoding the 0th level, this property implies that the equivalent channel seen by the bits is output-symmetric and that these bits are also uniformly distributed. In this case the probability of decoding error is the same for linear LDPC codes and LDPC coset codes, which allows us to use linear LDPC codes designed with the standard density evolution method~\cite{Richardson2001a}. This property no more holds when decoding the following levels, however recent results suggest that linear LDPC codes may still perform well with our coset coding scheme~\cite{Wang2005}. In order to further simplify the code design, we used irregular LDPC codes optimized for antipodal signaling over the AWGN channel as component codes. The block length used was 200,000 and graphs were randomly generated while avoiding cycles of length two and four. Despite this long block length, the perfomances of all constructed codes were still well below those of their ideal capacity achieving counterparts, and perfect error-correction can therefore only be achieved by using lower rates codes at each level. Cutting down the rate of all component codes would disclose far too many bits, however a careful choise of the code taking into account multiple iterations between levels make it possible to maintain a good efficiency.

\begin{figure}[htbp]
    \centering
        \includegraphics[width=10cm]{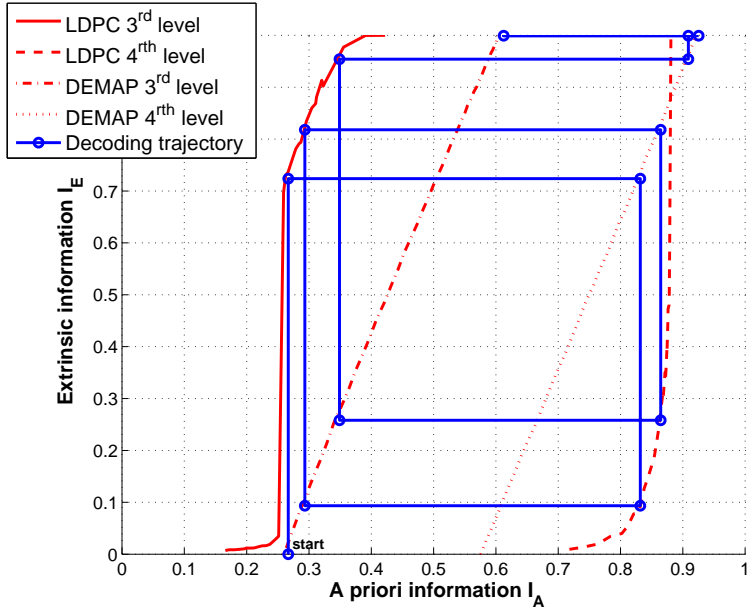}
    \caption{Iterative decoding trajectory averaged over 10 realizations.}
    \label{fig:trajectory}
\end{figure}

The practical code rate assignment is based on an analysis of the decoding process using EXIT charts~\cite{tenBrink2001}. Although there exist no theoretical results associated with EXIT charts for the Gaussian channel, they are a convenient tool to predict the exchange of information between the demappers and decoders involved in an iterative decoding scheme, based on how much extrinsic information ($I_E$) they compute from \textit{a priori} information ($I_A$). There is no closed-form expression of the EXIT curve $I_E=T_d(I_A)$ of the demapper characterized by Eq.~(\ref{eq:interlevel}) and of the LDPC EXIT curve $I_E=T_c(I_A)$ for $100$ iterations, however they can be obtained via Monte-Carlo simulations assuming Gaussian \textit{a priori} information~\cite{tenBrink2001}. Example of transfer curves are shown in Figure~\ref{fig:trajectory}. We observed that low rate codes gather extrinsic information at a slower pace than high-rate codes, therefore we decided to correct all errors 
by reducing the rate of the highest-rate code and by using iterations between levels to compensate for the poor performance of the lower rate code.

Let us now illustrate how code rates can be chosen on an example. We consider the situation where the SNR is 13~dB, for which the optimal constellation is given in table~\ref{tab:refconst}.  One would in theory need two ideal codes with rate 0.264 and 0.928. We used instead a code with rate 0.25 at the first level and looked for a high rate code that would gather enough extrinsic information to start the decoding process and correct all errors with an \textit{a priori} information of 0.928. As shown in Fig.~\ref{fig:trajectory}, a code with rate 0.86 was a good compromise. It is interesting to note that despite the approximations made in the computation of the EXIT curves, the real decoding trajectory is close to the expected behavior.

\subsection{Efficiency results}
The results obtained for various values of the noise variance are summarized in table~\ref{tab:res_sum}. For each SNR, the size of the constellation $\mathcal{X}$, the position of the constellation points and the probability distribution were optimized according to the procedure described in section~\ref{sec:Funda}, to ensure ${\left|I(X;Y_M)-C_M\right|}\leq 0.005$~bits while limiting the number of required codes to two.
\begin{table}[htbp]
    \centering
        \caption{Efficiency results.}
        \begin{tabular}{|c|c|c|c|c|c|c|c|}
            \hline
            $10\log_{10}\frac{1}{N_M}$   & $\left|\mathcal{X}\right|$    &   $I(X;Y_M)$ &    $C_M$ & $H(X)$  & Optimal rates &   Practical rates & Efficiency \\
            \hline
            \hline 2~dB & 4 &   0.684   &   0.685   & 1.603 & 0.189/0.891  & 0.16/0.86&90.9\%\\
            \hline 7~dB & 6 &   1.291   &   1.294   & 2.109 & 0.257/0.925/1& 0.24/0.86/1&90.9\%\\
            \hline 10~dB    & 8 &   1.726   &   1.730   & 2.502& 0.286/0.938/1 &    0.27/0.88/1& 95.71\%\\
            \hline 13~dB    & 12    &   2.192   &   2.194   & 3.000& 0.264/0.928/1/1 &  0.25/0.86/1/1& 96.15\%\\
            \hline 20~dB    & 28    &   3.327   &   3.329   & 4.149&    0.254/0.923/1/1/1 & 0.24/0.86/1/1/1& 97.6\%\\
            \hline
        \end{tabular}
    \label{tab:res_sum}
\end{table}
Let us point out that our method achieves good efficiency provided that two conditions are met. First, the constellation size required to maximize $I(X;Y_M)$ should be $\left|\mathcal{X}\right|\geq 4$ so that two LDPC codes can be used. Second, the codes rates required should not be too small so that we can construct good finite length codes. In practive this limited our simulations to situations where the SNR was above 2~dB.

\section{Opportunistic Security for Wireless Communications}
\label{sec:opportunSec}

This section describes an explicit secret key agreement protocol for wireless channels, exploiting the reconciliation algorithm described earlier. The proposed scheme closely follows the general approach presented in section~\ref{sec:Funda}, however all steps are modified to take into account the specific nature of the channels.

\subsection{System Setup}
\begin{figure}[htbp]
    \centering
        \includegraphics[width=10cm]{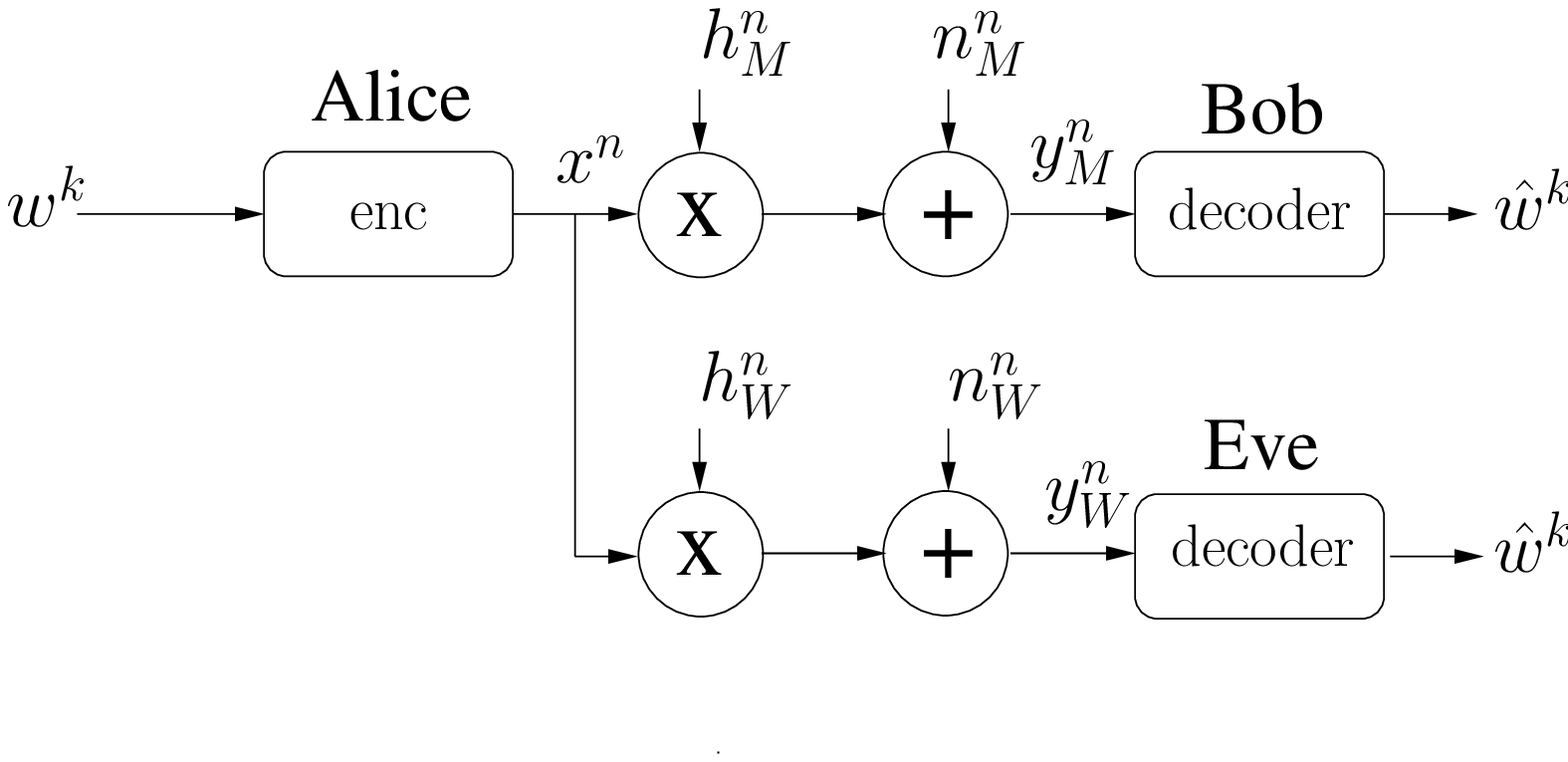}
    \caption{Wireless system setup.}
    \label{fig:setup}
\end{figure}

We consider the wireless system depicted in Fig.~\ref{fig:setup}.
Bob and Eve respectively observe the symbols sent by Alice through discrete-time Rayleigh-fading channels given by
\begin{eqnarray}
y_M(i) &=& h_M(i)x(i)+n_M(i),\\
y_W(i) &=& h_W(i)x(i)+n_W(i),\quad\mbox{subject to}\quad E\left\{\left|X(i)^2\right|\right\}\leq 1,
\end{eqnarray}
where $h_M(i)$ ($h_W(i)$) denotes the zero-mean fading coefficient of the main channel (wiretap channel), and $n_M(i)$ ($n_W(i)$) is a zero-mean complex Gaussian noise with variance $N_M$ ($N_W$). We further assume the fading coefficients and the noises to be independent, and the fading coefficients to remain constant over the transmission of several consecutive symbols (quasi-static fading). The instantaneous SNRs corresponding to a single realization $(h_M,h_W)$ of the fading coefficients are denoted $\gamma_M={\left|h_M\right|}^2/N_M$ and $\gamma_W={\left|h_W\right|}^2/N_W$, and the instantaneous capacities are then $C_M=\log_2(1+\gamma_M)$ and $C_W=\log_2(1+\gamma_W)$. As shown in~\cite{Barros2006}, the instantaneous secrecy capacity is
\begin{equation}
C_s = \left\{ \begin{array}{ll} C_M - C_W& \textrm{if $\gamma_M > \gamma_W$}\\
0 & \textrm{if $\gamma_M \leq \gamma_W$.}
\end{array} \right.
\label{eq:cs}
\end{equation}

\subsection{Secure Communication Protocol}
\begin{figure}[htbp]
    \centering
        \includegraphics[width=10cm]{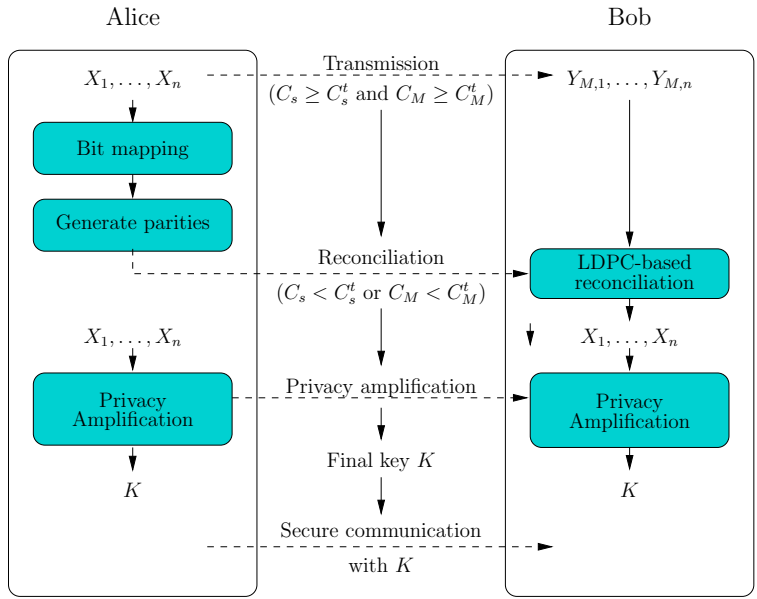}
    \caption{Flowchart of the opportunistic protocol.}
    \label{fig:flowchart}
\end{figure}

The fluctuations of the instantaneous secrecy capacity $C_s$ with time suggest the following opportunistic secret key agreement scheme (see also Fig.~\ref{fig:flowchart}).
\begin{itemize}
\item \textbf{Opportunistic transmissions.} When the estimated
instantaneous secrecy capacity $C_s$ and the instantaneous main channel capacity $C_M$
computed using the available CSI are greater than some thresholds $C_s^t\geq 0$ and $C_M^t\geq 0$,
Alice transmits random symbols at a rate equal to the capacity $C_M$ using a Gaussian shaped Quadrature
Amplitude Modulation scheme. We assume that Bob knows the channel fading coefficient $h_M$ and detects coherently the symbols sent by Alice, hence the fading channel can be viewed as two independent real Gaussian channels. The QAM constellation required to send close to $C_M$ bits/symbols can therefore be obtained by replicating the PAM scheme decribed in section~\ref{sec:Funda} in two dimensions. This phase is called "opportunistic transmission" since Alice and Bob take advantage of the channel realizations where they know they can exchange more information than Eve can intercept. The threshold $C_M^t$ is imposed by the reconciliation method which fails below a certain SNR, the choice of the threshold $C_s^t$ will be discussed in the next section.
\item \textbf{Reconciliation and privacy amplification.} When the estimated secrecy capacity or main channel capacity fall
below their respective theresholds, Alice and Bob extract a secure key from
the shared randomness  previously obtained. The reconciliation algorithm described in section~\ref{sec:reconciliation} allows Bob to recover Alice's symbols exactly, while limiting the additional information
sent over the channel. Privacy amplification with
universal hash functions is then used to distill a secret key,
taking into account all the information leaked to Eve during the
opportunistic transmission and reconciliation
stages.
\item \textbf{Secure communication.} Alice and Bob can finally use their secret key to
transmit messages, using either a one-time pad to ensure perfect
secrecy or any symmetric cypher.
\end{itemize}

Notice that the randomness sharing and privacy amplification steps rely at this point on a perfect estimation of the fading coefficients to calculate the instantaneous secrecy capacity and then correctly estimate the length of the secret key to distill. As we will see shortly, this assumption can be somewhat alleviated to consider a more realistic situation where only imperfect CSI (or a conservative estimate) is available for the wiretap channel.

\section{Results}

\subsection{Performance Measures for Information-Theoretically Secure Communications}
The performance of the opportunitic protocol in the case of perfect channel state information will be evaluated with the following two measures: {\em average $\eta$-secure throughput $\overline{\mathcal{T}_s}(\eta)$  and the average $\eta$-communication throughput $\overline{\mathcal{T}_c}(\eta)$ }. The average $\eta$-secure throughput $\overline{\mathcal{T}_s}(\eta)$ is defined as the average number of secured message bits transmitted per channel use, where $\eta$ is the ratio of secret-key bits used per message bits ($\eta\leq 1$). Note that in a secret key agreement scenario, the secret-key generation rate does not contribute to the $\eta$-secured throughput since the key itself does not convey any information. When $\eta=1$, $\overline{\mathcal{T}_s}(\eta)$ corresponds to a perfectly secure communication obtained with a one-time pad encryption, whereas $\overline{\mathcal{T}_s}(\eta)$ for $\eta<1$ only represents an encrypted message rate with secret keys. If $k_s$ is the key length required for encryption, the corresponding key renewal rate is $k_s/\eta$. Similarily the average $\eta$-communication throughput $\overline{\mathcal{T}_c}(\eta)$ is defined as the average number of non-secure message bits transmitted per channel use. In the case of secret-key agreement, the communication rate used for reconciliation and privacy amplification has to be deduced from the total communication throughput.


Let us now evaluate $\overline{\mathcal{T}_s}(\eta)$ for our protocol.  Let $\mathcal{D}=\left\{(\gamma_M,\gamma_W):C_s\geq
C_s^t,C_M>C_M^t\right\}$ be the set of fading realizations for which an oportunistic transmission is performed and let $\overline{\mathcal{D}}$ denote its complement in $\mathbb{R}_+^2$. For a given random variable X depending on the fading realization, let $\avg{X}{\mathcal{D}}$ denote its average over all fading realizations in $\mathcal{D}$. We will assume that fading coefficients remain constant over the transmission of $n\gg 2s+2+r_0$ symbols, where $s$ and $r_0$ are the safety parameters used during privacy amplification. We can then the neglect the penalty inflicted by privacy amplification and assume that the opportunistic transmissions provide on average
\begin{equation}
\avg{\beta I(X;Y_W)-I(X;Y_W)}{\mathcal{D}} \approx \avg{\beta C_M-C_W}{\mathcal{D}}
\end{equation}
secret key bits per symbol transmitted, which can then be used to secure
\begin{equation}
\label{eq:securethroughput}
\overline{\mathcal{T}_s}(\eta) = \eta^{-1} \avg{\beta C_M-C_W}{\mathcal{D}}
\end{equation}
bits of message per symbol. From section~\ref{sec:reconciliation}, we know that reconciliation requires the transmission of $\avg{H(X|Y_M)+(1-\beta) I(X;Y_M)}{\mathcal{D}}\approx\avg{H(X)-\beta C_M}{\mathcal{D}}$ additional bits per symbols on average. The minimum size of a universal familiy of hash functions $\mathcal{G}:{\left\{0,1\right\}}^{n_rec}\rightarrow{\left\{0,1\right\}}^{k}$ is at least $2^{n_{rec}-k}$~\cite{Stinson1991} and privacy amplication therefore requires the transmission of $n_{rec}-k$ bits. No hashing scheme is known to achieve this bound for any $n_{rec}$, therefore we will consider the more realistic situation where privacy amplification requires the transmission of $n_{rec}$ bits. For instance, this can be achieved with the following family~\cite{Wegman1981}:
\begin{equation}
\mathcal{H}_{\mbox{GF}(2^{n_{rec}})\rightarrow\{0,1\}^{n_{key}}} = \left\{h_c: c\in\mbox{GF}(2^{n_{rec}})\right\},
\end{equation}
where $h_c(x)$ is defined as $n_{key}$ distinct bits of the product $cx$ in a polynomial representation of $\mbox{GF}(2^{n_{rec}})$. Finally, since the maximum number of non-secure bits transmitted is at most the capacity of the main channel, we obtain:
\begin{equation}
\label{eq:commthroughput}
\overline{\mathcal{T}_c}(\eta) = \avg{C_M}{\overline{\mathcal{D}}} - \avg{H(X)-\beta C_M}{\mathcal{D}} -  \avg{H(X)}{\mathcal{D}}-\eta^{-1} \avg{\beta C_M-C_W}{\mathcal{D}}.
\end{equation}
Notice that $\overline{\mathcal{T}_c}(\eta)$ may be negative when $\prob(C_s\geq C_s^t)\gg \prob(C_s < C_t^s)$. This situation corresponds to a regime where Alice and Bob generate keys faster than they use them, which can be avoided by adjusting the parameter $C_t$ so that $\overline{\mathcal{T}_c}(\eta)$ remains positive. In the remaining of the paper, we will be interested in the ultimate performance of the protocol, therefore according to section~\ref{sec:reconciliation} we will assume $H(X)\approx C_M+2$ but unless otherwise specified we will use $C_M^t=0$ and $\beta=1$.

\begin{figure}[htbp]
    \centering
        \includegraphics[width=10cm]{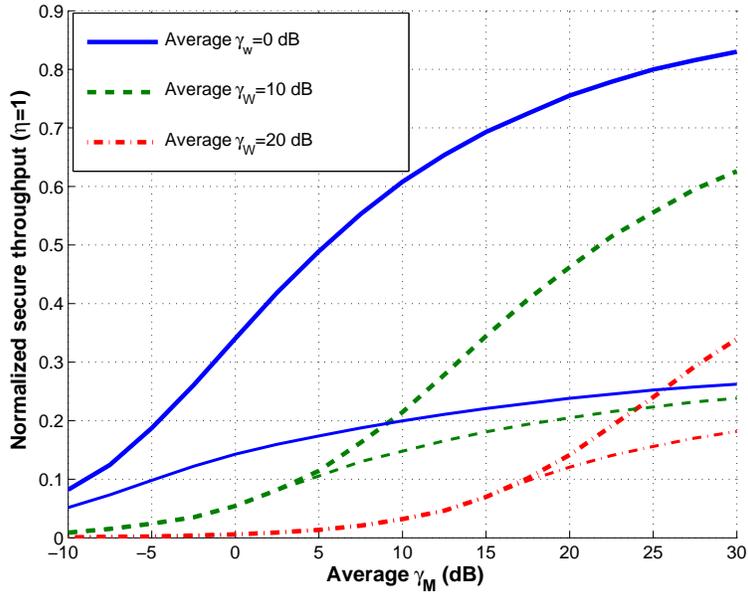}
    \caption{Average secure throughput (thin lines) and average secrecy capacity (thick lines). All throughputs are normalised to the channel capacity of a Gaussian channel with same average SNR $\overline{\gamma}_M$.}
    \label{fig:throughput_vs_capacity}
\end{figure}

The maximum average secure throughput for $\eta=1$ achievable by the opportunistic protocol is shown Fig.~\ref{fig:throughput_vs_capacity}.
As expected the protocol is in general sub-optimal since most of the main channel capacity has to be sacrificed for key agreement. Interestingly when the wiretap channel average SNR $\overline{\gamma}_W$ is well above the main channel average SNR $\overline{\gamma}_M$, all the additional communication required for reconciliation and privacy amplification as well as the communication secured by a one-time pad, can be performed when the secrecy capacity is zero. In this case, the protocol incurs no loss of secure communication rate.

\begin{figure}[htbp]
    \centering
        \includegraphics[width=10cm]{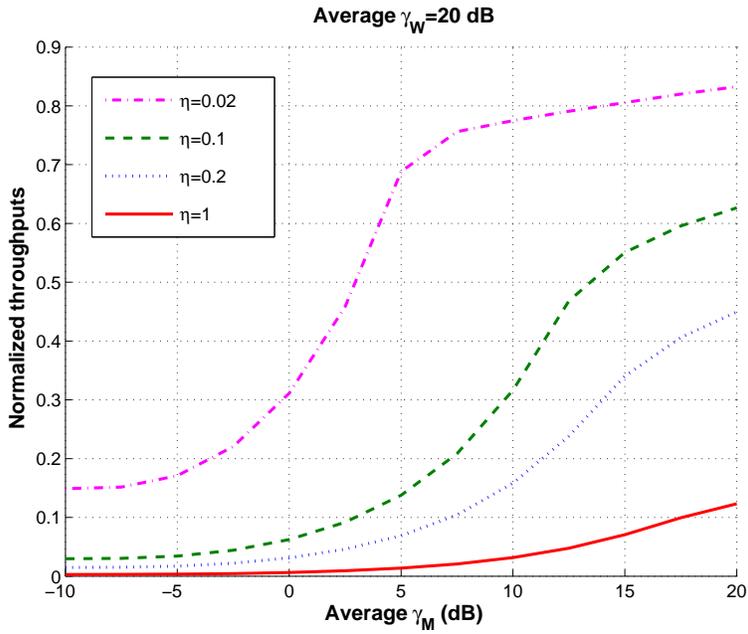}
    \caption{Secure throughput for various values of $\eta$.}
    \label{fig:eta_throughputs}
\end{figure}

Fig.~\ref{fig:eta_throughputs} shows the secure throughputs obtained for different values of $\eta$. Strictly speaking, the protocol does not provide any information theoretic security in this regime, since the keys generated are used to encode several bits.
Nevertheless, this result shows that the protocol provides an efficient and potentially fast way of exchanging information-theoretically secure keys. In this mode of operation, it could be tailored with standard secure encryption algorithms (such as the AES with 192 bits) to strenghten the current level of security of wireless communications.

\subsection{Mitigating the Effects of Imperfect CSI}

Let us now briefly discuss the impact of imperfect channel state information. We can reasonably assume that Bob cooperates with Alice, which allows her to obtain a perfect estimate of the main channel fading coefficient. Unfortunately Eve may not be as helpful and Alice's knowledge of the wiretap fading is more likely to be noisy. In order to assess the performance of our protocol under more realistic conditions, we model Alice's estimate of Eve's fading coefficient by $\hat{h}_W = h_W+n_W$, where $h_W$ is the true fading coefficient and $n_W$ is a zero-mean complex Gaussian noise with known variance $\sigma^2$ per dimension.
If Alice applies the previous protocol blindly, her estimation $\hat{C}_s$ of the instantaneous secrecy capacity will generally differ from the real secrecy capacity $C_s$. The situations where $\hat{C}_s\leq C_s$ do not impact the secrecy of the key agreement, however when $\hat{C}_s>C_s$, Alice understimates the information leaked to the eavesdropper and subsequently generate keys whose entropy is not maximum. Let $\hat{K}$ denote the $\hat{k}$-bit key generated by Alice based on her estimation $\hat{C}_s$. From theorem~\ref{th:general_privacy_amp}, the uncertainty on $\hat{K}$ of the eavesdropper is bounded as follows:
\begin{equation}
\hat{k}\geq H(\hat{k}|G,E=e) \geq \hat{k} -\frac{2^{n(C_W-\hat{C}_W)-r_0}}{\ln 2} = \hat{k} -\frac{2^{n(C_W-\hat{C}_W-\alpha)}}{\ln 2},
\end{equation}
where we have introduced the parameter $\alpha=r_0/n$. As long as $C_W-\hat{C}_W\leq\alpha$, the uncertainty of the key $\hat{K}$ lies within 1.5 bits of its maximal value and can be regarded as secret, however when $C_W-\hat{C}_W>\alpha$ the lower bound on $H(\hat{K}|G,E=e)$ decreases exponentially in the difference $C_W-\hat{C}_W-\alpha$.

\begin{figure}[htbp]
    \centering
        \includegraphics[width=10cm]{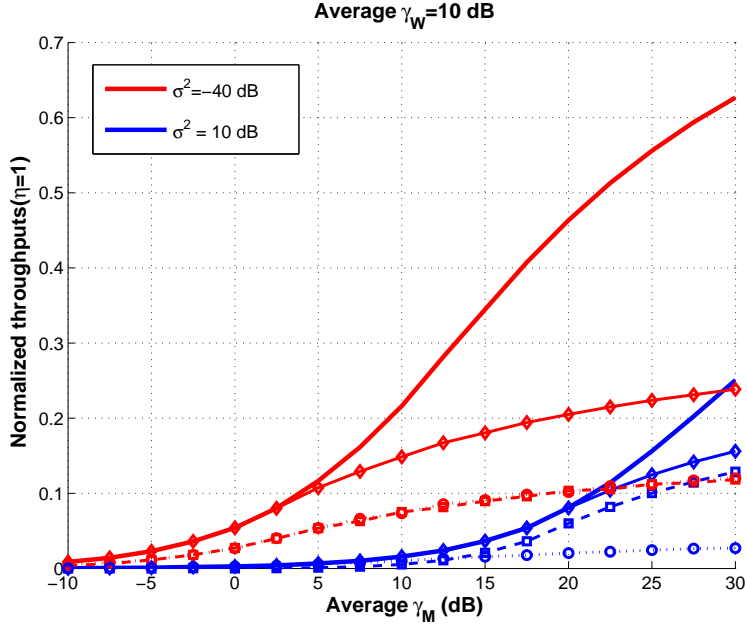}
        \caption{Impact of imperfect CSI. Thicker lines represent the estimated average secrecy capacity. The diamond lines ($\diamond$) represent Alice's targeted average secure throughput with her imperfect CSI, the square lines ($\boxdot$) and circle lines ($\circ$) respectively represent the true average secure throughput and average leaked throughput. All throughputs are normalised to the channel capacity of a Gaussian channel with same average SNR $\overline{\gamma}_M$.}
    \label{fig:CSI_0}
\end{figure}

The introduction of imperfect CSI and the use of the parameter $\alpha$ slightly modify the expression of the average secure and communication given by Eq.~(\ref{eq:securethroughput}) and~(\ref{eq:commthroughput}). Let $\mathcal{D}=\left\{(\hat{\gamma_M},\gamma_W):\hat{C_s}\geq
C_s^t\right\}$, then
\begin{eqnarray}
\overline{\mathcal{T}_s}(\eta) &=& \eta^{-1} \avg{\hat{C_s}-\alpha}{\mathcal{D}}\\
\overline{\mathcal{T}_c}(\eta) &=& \avg{C_M}{\overline{\mathcal{D}}} - \avg{H(X)-C_M}{\mathcal{D}} -  \avg{H(X)}{\mathcal{D}}-\eta^{-1} \avg{\hat{C}_s-\alpha}{\mathcal{D}}.
\end{eqnarray}
The threshold $C_s^t\geq\alpha$ should once more be chosen such that $\overline{\mathcal{T}_c}(\eta)\geq 0$. Contrary to the situation where perfect CSI is available, the average secure throughput defined above is not sufficient to characterize the security of the system. In fact it only represents Alice's targeted secure communication rate, which might be different from the true secure communication rate. Hence we need to introduce the true average secure throughput $\overline{\mathcal{R}}_s$ and the average leaked throughput $\overline{\mathcal{R}}_l$ defined as:
\begin{eqnarray}
\overline{\mathcal{R}}_s &=& \eta^{-1} \avg{\hat{C_s}-\alpha}{\mathcal{D}_s},\\
\overline{\mathcal{R}}_c &=& \eta^{-1} \avg{\hat{C_s}-\alpha}{\mathcal{D}_l},
\end{eqnarray}
where $\mathcal{D}_s=\left\{(\hat{\gamma_M},\gamma_W):\hat{C_s}\geq
C_s^t,\hat{C}_s-C_s \leq\alpha\right\}$ and $\mathcal{D}_l=\left\{(\hat{\gamma_M},\gamma_W):\hat{C_s}\geq
C_s^t,\hat{C}_s-C_s >\alpha\right\}$.
These expressions cannot be computed in close form but can be obtained with Monte-Carlo simulations. Figure~\ref{fig:CSI_0} shows the results obtained for an estimation noise variance of $\sigma^2=10$ and  $\sigma^2=0.0001$ when $\eta=1$ and $\alpha=0$ (i.e. the safety parameter $r_0\ll n$).

\begin{figure}[htbp]
    \centering
        \includegraphics[width=10cm]{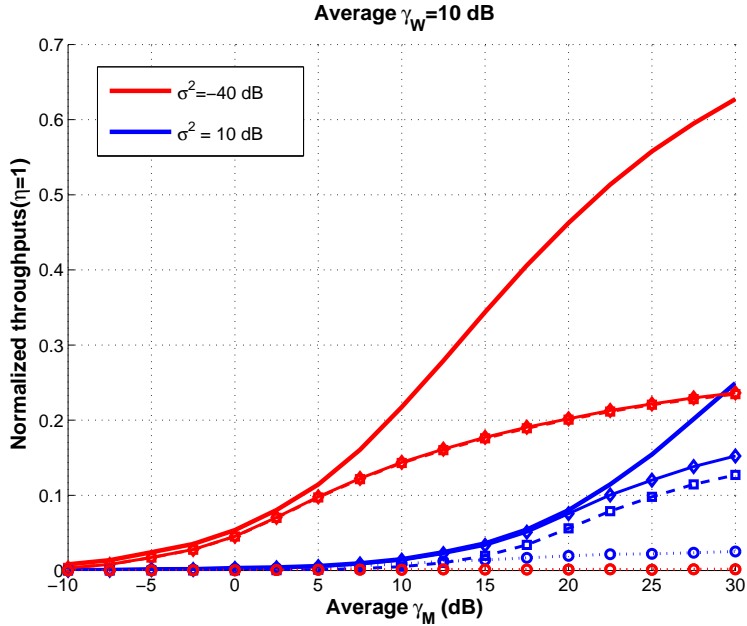}
        \caption{Mitigation of imperfect CSI. Thicker lines represent the estimated average secrecy capacity. The diamond lines ($\diamond$) represent Alice's targeted average secure throughput with her imperfect CSI, the square lines ($\boxdot$) and circle lines ($\circ$) respectively represent the true average secure throughput and average leaked throughput. All throughputs are normalised to the channel capacity of a Gaussian channel with same average SNR $\overline{\gamma}_M$.}
    \label{fig:CSI_0.1}
\end{figure}

Interestingly when Alice has a bad estimation of the wiretap channel fading coefficient and if the main channel SNR is well above the wiretap channel SNR, most of the keys generated are still secret. This unexpected behavior can be explained by the asymmetry of the distribution $p(\hat{\gamma}_W|\gamma_W)$ which forces Alice to undersestimate $C_W$ most of the time. On the other hand when her estimations of the wiretap CSI improves, she becomes equally likely to overstimate or understimate $C_W$, therefore  $\overline{\mathcal{R}}_c\approx\overline{\mathcal{R}}_s$ and half of the keys generated are then insecure. The impact of imperfect of imperfect channel state information can be somewhat mitigated by increasing the parameter $\alpha$. In fact, $\alpha>0$ plays the role of a safety margin and reduces the length of the generated keys. By increasing $\alpha$, the average leaked throughput can be made arbitrarily small, but this also decreases the achievable secure throughput. Figure~\ref{fig:CSI_0.1} shows the results obtained for $\alpha=0.1$.
When $\sigma^2=0.0001$, the secure throughput loss is negligible, however this slight increase in $\alpha$ suffices to ensure the secrecy of the keys generated. The mitigation is less effective when $\sigma^2=10$, and a further increase of $\alpha$ would be necessary to reduce the leaked throughput.







\section{Conclusions and Future Work}
In the second of this two-part paper on wireless information-theoretic security,
we proposed a protocol based on one-way communications providing secure communication over quasi-static wireless channels. This scheme opportunistically exploits the fluctuations of the fading coefficients to exchange information-theoretially secure keys, which are then be used to encrypt messages. We analysed the security provided by the protocol in the idealized case where the channel state information of the wiretap channel is known, but also showed that secure communication is still achievable in the more realistic situation where only imperfect channel state information is available. The fundamental security limits in both scenarios were studied in Part I.

The performance and complexity of the proposed scheme mainly rely on those of the reconciliation algorithm. Our LDPC-based reconciliation method is near-optimal over a wide range of signal-to noise ratios, however the memory requirements and the complexity may still be too high for embedded or low-cost systems. In future work, we will investigate new code constructions to in order to reduce the hardware requirements while still maintaining the same level of performance.

Let us finally mention that even though the encryption used in our scheme could be performed with a one-time pad to ensure perfect security, the protocol may be of higher interest if tailored with existing secret-key encryption methods (e.g. DES, AES) to strenghten their current level of security.

\bibliographystyle{IEEEtran}

\end{document}